\documentstyle[preprint,tighten,aps]{revtex}
\begin{document}
\draft

\title
{Renormalized sextic coupling constant \\ 
for the two-dimensional Ising model \\ 
from field theory} 

\author{A.~I.~Sokolov and E.~V.~Orlov}

\address
{Department of Physical Electronics, \\ 
Saint Petersburg Electrotechnical University, \\
Professor Popov Street 5, St. Petersburg, 197376, Russia}

\maketitle

\begin{abstract}
The field-theoretical renormalization group approach is used to 
estimate the universal critical value $g_6^*$ of renormalized sextic 
coupling constant for the two-dimensional Ising model. A four-loop 
perturbative expansion for $g_6$ is calculated and resummed by means 
of the Pad\'e-Borel-Leroy technique. Under the optimal value 
of the shift parameter $b$, providing the fastest convergence of the 
iteration procedure the estimates $g_6^* = 1.10$, $g_6^*/{g_4^*}^2 = 2.94$
are obtained, which agree quite well with those deduced recently by Zinn, 
Lai, and Fisher [Phys. Rev. E {\bf 54}, 1176 (1996)] from the 
high-temperature expansions.

\vspace{0.3cm}

PACS numbers: 05.70.Jk, 11.10.Gh, 64.60.Ak, 64.60.Fr
 
\end{abstract}

\newpage

\section{Introduction}
\label{sec:1}

The two-dimensional (2D) Ising model in a zero magnetic field  
was exactly solved by L.~Onsager more than 50 years ago. Since then, 
many attempts were made to develope a theory describing the behavior 
of this model under an external field but, nevertheless, an exact 
analytical description of its magnetized state has not been given up 
to now. On the other hand, for decades the critical behavior of various 
systems in ordering fields has attracted constant attention, being of prime 
interest both for theorists and experimentalists. Recently, 
the free energy (effective action) and, in particular, higher--order 
renormalized coupling constants $g_{2k}$ for the basic models of 
phase transitions, have become the target of intensive theoretical studies 
\cite{BB,TW,Ts94,R,S96,ZLF,SUO,SOU,GZ,Morr,BC97,Ts97,S98,PV}. 
These constants are related to the non--linear susceptibilities 
$\chi_{2k}$, which enter the scaling equation of state and thus play very 
important role at criticality. Along with critical exponents and 
critical amplitude ratios, they are universal, i.e., possess, under 
$T \to T_c$, numerical values that are not sensitive to the physical 
nature of phase transition, depending only on the system dimensionality 
and the symmetry of the order parameter. 

Calculation of the universal critical values of $g_6$, $g_8$, etc. 
for three-dimensional Ising and $O(n)$-symmetric models by  
a number of analytical and numerical methods showed that the 
field-theoretical renormalization group (RG) approach in fixed 
dimensions yields most accurate numerical estimates for these 
quantities. It is a consequence of a rapid convergence of the 
iteration schemes originating from renormalized perturbation 
theory. Indeed, the resummation of four- and five-loop RG 
expansions by means of the Borel-transformation-based procedures
gave the values for $g_6^*$, which differ from each other by less
than $0.5 \%$ \cite{SOU,GZ} while the use of resummed three-loop RG 
expansion enabled one to achieve an apparent accuracy no worse than 
$1.6 \%$ \cite{SUO,S98}. 

The field-theoretical RG approach, being very effective in 3D 
(see, e.g., Refs.~\cite{BNM,LGZ,AS,S98}) is also known to be powerful 
in two dimensions. Properly resummed four-loop RG 
expansions lead to fair numerical estimates for critical exponents 
of a 2D Ising model \cite{BNM} and give reasonable results for its 
random counterpart \cite{MSS,M}. It is natural, therefore, to use 
the field theory for the calculation of renormalized higher-order coupling 
constants of a 2D Ising model. In this paper, the 2D RG expansion for 
renormalized coupling constant $g_6$ will be calculated, and the numerical 
estimate for its universal critical value $g_6^*$ will be obtained.

\section{RG expansion for sextic coupling constant}
\label{sec:2}

Within the field-theoretical language, the two-dimensional Ising 
model in the critical region is described by Euclidean scalar field 
theory with the Hamiltonian
\begin{equation}
H = 
\int d^2 x \Biggl[ {1 \over 2} m_0^2 {\varphi}^2
 + {1 \over 2} (\nabla \varphi)^2
+ \lambda {\varphi}^4 \Biggr] \  , 
\label{eq:1}
\end{equation}
where a bare mass squared $m_0^2$ is proportional to 
$T - T_c^{(0)}$, $T_c^{(0)}$ being the phase transition
temperature in the absence of the order-parameter fluctuations. 
Taking fluctuations into account results in renormalizations
of the mass $m_0 \to m$, the field $\varphi = \varphi_R \sqrt{Z}$, 
and the coupling constant $\lambda = m^2 Z_4 Z^{-2} g_4$. Moreover,
thermal fluctuations give rise to many-point correlations 
$<\varphi(x_1) \varphi(x_2)...\varphi(x_{2k})>$ and, 
correspondingly, to higher-order terms in the expansion of the 
free energy in powers of magnetization $M$:
\begin{equation}
F(M) - F(0) = m^2 \Biggl[{1 \over 2}{M^2 \over Z} + 
g_4 \Biggl({M^2 \over Z} \Biggr)^2 +
\sum_{k=3}^{ \infty} {g_{2k} \Biggl({M^2 \over Z} \Biggr)^k} \Biggr].
\label{eq:2}
\end{equation}
In the critical region, where fluctuations are so strong that 
they completely screen out the initial (bare) interaction, the 
behavior of the system becomes universal and dimensionless 
effective couplings $g_{2k}$ approach their asymptotic limits 
$g_{2k}^*$.

In order to estimate $g_6^*$, we will calculate RG expansion 
for $g_6$ and then apply Pad\'e-Borel-Leroy resummation 
technique to get proper numerical results. Accurate enough 
numerical estimates, as is well known, may be extracted only 
from a sufficiently long RG series. Below we will find the expression
for $g_6$ in the four-loop approximation that will be shown to 
provide an interesting amount of fair numerical estimates.

The method of calculating the RG series that we use here is 
straightforward. Since in two dimensions higher-order bare 
couplings are believed to be irrelevant in RG sense, renormalized
perturbative series for $g_6$ can be obtained from a conventional
Feynman graph expansion of this quantity in terms of the only 
bare coupling constant - $\lambda$. In its turn, $\lambda$ 
may be expressed perturbatively as a function of renormalized 
dimensionless quartic coupling constant $g_4$. Substituting 
corresponding power series for $\lambda$ into original expansion, 
we can obtain the RG series for $g_6$. As was recently shown 
\cite{S96,SUO,SOU} the one-, two-,  three- and four-loop 
contributions are formed by 1, 3, 16, and 94 one-particle 
irreducible Feynman graphs, respectively. Their calculation gives:
\begin{equation}
g_6 = {36 \over \pi}{{\Biggl({\lambda Z^2 \over m^2} \Biggr)}^3}{\Biggl
[ 1 - 11.817855\ {\lambda Z^2 \over m^2} + 
110.37270\ {\Biggl({\lambda Z^2 \over m^2} \Biggr)^2} -
985.575\ {\Biggl({\lambda Z^2 \over m^2} \Biggr)^3} \Biggr]} ,
\label{eq:3} 
\end{equation}
The perturbative expansion for $\lambda$ emerges directly from the 
normalizing condition $\lambda = m^2 Z_4 Z^{-2} g_4$ and the 
well-known expansion for $Z_4$ \cite{BNM}:
\begin{equation}
Z_4 = 1 + {9 \over \pi}\ g_4 + 5.1288114\ g_4^2 + 
10.511670\ g_4^3 + O(g_4^4).
\label{eq:4}
\end{equation} 
Combining these expressions we obtain
\begin{equation}
g_6 = {36 \over \pi}\ {g_4^3}\ {\Bigl( 1 - 3.2234882\ g_4 + 
14.957539\ g_4^2 - 85.7810\ g_4^3 \Bigr)}.
\label{eq:5}
\end{equation}
This series will be used for estimation of the universal number $g_6^*$.
  
\section{Resummation and numerical estimates}
\label{sec:3}

Being a field-theoretical perturbative expansion the series 
(\ref{eq:5}) has factorially growing coefficients, i.e., it is  
divergent (asymptotic). Hence, direct substitution of the fixed 
point value $g_4^*$ into (\ref{eq:5}) would not lead to satisfactory 
results. To get reasonable numerical estimate for $g_6^*$, some 
procedure making this expansion convergent should be applied. 
As is well known, the Borel-Leroy transformation 
\begin{equation}
f(x) =  \sum_{i = 0}^{\infty} c_i x^i = 
\int\limits_0^{\infty} t^b e^{-t} F(xt) dt , \qquad
F(y) = \sum_{i = 0}^{\infty} {c_i \over (i+b)!} y^i .
\label{eq:6}
\end{equation}
diminishing the coefficients by the factor $(i+b)!$, can play a role 
of such a procedure. Since the RG series considered turns out to be 
alternating, the analytical continuation of the Borel transform may 
be then performed by using Pad\'e approximants. With the 
four-loop expansion (\ref{eq:5}) in hand, we can construct three 
different Pad\'e approximants: [2/1], [1/2], and [0/3]. To 
obtain proper approximation schemes, however, only diagonal [L/L] 
and near--diagonal Pad\'e approximants should be employed 
\cite{BGM}. That's why further we limit ourselves to approximants 
[2/1] and [1/2]. Moreover, the diagonal Pad\'e approximant 
[1/1] will be also dealt with, although this corresponds, in fact, to 
the usage of the lower-order, three-loop RG approximation. 

The algorithm of estimating $g_6^*$ that we use here is as follows. 
Since the Taylor expansion for the free energy contains as 
coefficients the ratios $R_{2k} = g_{2k}/g_4^{k-1}$, rather than 
the renormalized coupling constants themselves,
\begin{equation}
F(z) - F(0) = {m^2 \over g_4} \Biggl({z^2 \over 2} + z^4 + R_6 z^6 
+ R_8 z^8 + ... \Biggr), \qquad \qquad z^2 = {g_4 M^2 \over Z},  
\label{eq:7}
\end{equation}
we work with the RG series for $R_6$. It is resummed in three 
different ways based on the Borel-Leroy transformation and the 
Pad\'e approximants just mentioned. The Borel-Leroy integral is 
evaluated as a function of the parameter $b$ under $g_4 = g_4^*$. 
For the fixed point coordinate the value, $g_4^* = 0.6125$ 
(Refs. \cite{ZLF,B,BC96}) is adopted, which was extracted from lengthy 
high-temperature expansions and is believed to be the most 
accurate estimate for $g_4^*$ available today. The optimal value 
of $b$, providing the fastest convergence of the iteration scheme, is 
then determined. It is deduced from the condition that the Pad\'e 
approximants employed should give, for $b = b_{opt}$, 
the values of $R_6^*$, which are as close as possible to each other. 
Finally, the average over three estimates for $R_6^*$ is found and 
claimed to be a numerical value of this universal ratio. 
 
The results of our calculations are presented in Table 1. 
As one can see, for $b = 1.24$ all three working approximants 
lead to practically identical values of $R_6^*$. Hence,
we conclude that for 2D Ising model at criticality   
\begin{equation}
R_6^* = 2.94, ~~~~~~~g_6^* = 1.10. \ \
\label{eq:8}
\end{equation}
How close to their exact counterparts may these numbers be?  
To clear up this point, let us discuss the sensitivity of numerical 
estimates given by RG expansion (\ref{eq:5}) to the type of 
resummation. The content of Table 1 implies that, among others, 
the results given by Pad\'e approximant $[2/1]$ turn out to 
be most strongly dependent on the parameter $b$. This situation 
resembles that of 3D $O(n)$-symmetric model where Pad\'e 
approximants of $[L-1/1]$ type for the $\beta$-function and critical 
exponents lead to numerical estimates demonstrating 
appreciable variation with $b$, while for diagonal and near-diagonal 
approximants the dependence of the results on the shift parameter 
is practically absent \cite{S98,BNM,AS}. In our case, Pad\'e 
approximants $[1/1]$ and $[1/2]$ may be referred to as generating such 
"stable" approximations for $g_6^*$. Since for $b$, varying from 0 
to 15 (i.e., for any reasonable $b$), the magnitude of $g_6^*$ averaged 
over these two approximations remains within the segment (1.044, 1.142), 
it is hardly believed that the values (\ref{eq:8}) can differ from 
the exact ones by more than 5$\%$.   

Another way we propose to judge how accurate our numerical results 
are is based on the comparison of the values of $g_6^*$ given by four 
subsequent RG approximations available. While within the one-loop 
order we get $g_6^* = 2.633$, which is obviously very bad estimate, 
taking into account of higher-order RG contributions 
to $g_6$ improves the situation markedly. Indeed,  two-, three-, 
and four-loop RG series, when resummed by means of the 
Pad\'e-Borel technique with use of "most stable" approximants 
$[0/1]$, $[1/1]$, and $[1/2]$ yield for $g_6^*$ the values 0.981, 
1.129, and 1.051, respectively. Since this set of numbers demonstrates 
an oscillatory convergence, one may expect that the exact value of 
renormalized sextic coupling constant lies between the higher-order
three-loop and four-loop estimates. It means that the deviation 
of numbers (\ref{eq:8}) from the exact values would not exceed 5$\%$.  

Thus, we see that the four-loop RG expansion for $g_6$ and elaborated 
approximation scheme lead to accurate enough numerical data. On the 
other hand, the above arguments, as is always the case when we deal 
with diverging 
series, should be interpreted as suggestive, i.e., they would help us 
to fix only an apparent accuracy. It is of prime importance therefore 
to compare our estimates with those obtained by other methods. 
Recently, S.-Y. Zinn, S.-N. Lai, and M. E. Fisher, analyzing the high 
temperature series for various 2D Ising lattices, found
that $R_6^* = 2.943 \pm 0.007$ \cite{ZLF}; an almost identical value 
was obtained later in Ref.~\cite{PV}. Our result for $R_6^*$ is seen
to be in a brilliant agreement with this number. Of course, the practical 
coincidence of the lattice and four-loop RG estimates is occasional, 
and cannot be considered as a manifestation of extremely high accuracy 
of the methods discussed. The closeness of these estimates to each other, 
however, unambiguously demonstrates the power of both approaches. 
Moreover, such a closeness shed light on the role of a singular 
contribution to $g_6$, which can not be found perturbatively: 
this contribution is seen to be numerically small.   

It is instructive also to compare our results with those given by another 
field-theoretical approach - the famous $\epsilon$-expansion. Today, 
for the Ising systems, only three terms in the $\epsilon$--expansion for 
$R_6$ are known \cite{ZJ}:
\begin{equation}
R_6^* = 2 \epsilon \Biggl(1 - {10 \over {27}}\ \epsilon + 
0.63795\ {\epsilon}^2 \Biggr). 
\label{eq:9}
\end{equation}   
Let us apply a simple Pad\'e-Borel procedure to this series 
as a whole and to the series in brackets, and then let $\epsilon = 2$. 
We find $R_6^* = 3.19$ and $R_6^* = 3.12$, respectively, i.e., the numbers 
differ from our estimate by less that 9$\%$. Keeping in mind the lack 
of a small parameter, these values of $R_6^*$ may be referred to as 
consistent. We believe that the proper account for higher-order terms 
in the $\epsilon$-expansion for $R_6$ will make corresponding numerical 
estimates closer to those extracted from 2D RG and high-temperature 
series. 

\section{Conclusion}    
\label{sec:4}
    
To summarize, we have calculated the four-loop RG expansion for the 
renormalized sextic coupling constant $g_6$ of the two-dimensional 
Ising model. Resummation of this series by the 
Pad\'e-Borel-Leroy method has lead at criticality, under the 
optimal value of the parameter $b$, to the results ~$g_6^* = 1.10$, 
~$g_6^*/{g_4^*}^2 = 2.94$. Having analyzed the sensitivity of the RG 
estimates to the type of resummation procedure, and the character of 
their dependence on the order of the RG approximation, an apparent 
accuracy of these values has been argued to be no worse than 5$\%$. 
Comparison of the RG estimates with their counterparts given by other 
approaches has shown that they are in very good agreement with 
those deduced recently for 2D Ising lattices from high-temperature 
expansions, and consistent as well with the result given by the 
Pad\'e-Borel resummed three-loop $\epsilon$-expansion for $R_6^*$.

{\rm Notes added in proof}. (i) Apart from Refs. 1-14, the higher order 
coupling constants in $D$ dimensions were studied in Refs. \cite{24,25,26}. 
We are grateful to Dr. S. Boettcher, who brought these papers to our 
attention. (ii) Very good agreement exists between the first number 
Eq. (8) and the estimate $R_6^* = 2.95 \pm 0.03$ (Ref. 14) obtained 
recently by matching the $\epsilon$-expansion available with the exact 
results known for $D = 1$ and $D = 0$. It may be considered as an extra 
argument in favor of the high effectiveness of the field theory in the 
problem duscussed.   

\acknowledgments 
 
We thank B. N. Shalaev for interesting discussions and E. Vicari for 
sending the paper (Ref. 14). This work was supported by the Ministry 
of General and Professional Education of the Russian Federation under 
Grant No. 97-14.2-16.

\widetext
\begin{table}
\caption{The values of $R_6^*$ obtained by means of the 
Pad\'e-Borel-Leroy technique for various $b$ within 
three-loop (approximant $[1/1]$) and four-loop (approximants 
$[1/2]$ and $[2/1]$) RG approximations. The estimate for $b = 1$ 
in the middle line is absent because corresponding Pad\'e
approximant turnes out to be spoilt by a positive axis pole.}
\begin{tabular}{ccccccccccc}
$b$ & 0 & 1 & 1.24 & 2 & 3 & 4 & 5 & 7 & 10 & 15 \\
\tableline
$[1/1]$ & 2.741 & 2.908 & 2.937 & 3.009 & 3.077 & 3.125 & 3.161 
& 3.212 & 3.258 & 3.301 \\
\tableline
$[1/2]$ & 2.827 &  -  & 2.936 & 2.877 & 2.853 & 2.838 & 2.828 
& 2.814 & 2.800 & 2.787 \\
\tableline
$[2/1]$ & 3.270 & 2.988 & 2.936 & 2.800 & 2.667 & 2.568 & 2.491 
& 2.380 & 2.273 & 2.171 \\
\end{tabular}
\label{table1}
\end{table}

\end{document}